\documentclass[twocolumn]{aa}
\usepackage{graphicx}
\usepackage{epsfig}
\usepackage{rotating}
\begin{document}
   \title{Detection of the H92$\alpha$ Recombination Line from NGC 4945}

   \titlerunning{Recombination Lines from NGC 4945}

   \author{A.L. Roy\inst{1,2,3,4}, 
           T. Oosterloo\inst{5},
           W.M. Goss\inst{2}, \and
           K.R. Anantharamaiah\inst{3}\fnmsep\thanks{deceased}
          }

   \authorrunning{A. L. Roy et al.}

   \institute{$^{1}$Max-Planck-Institut f\"ur Radioastronomie, Auf dem
              H\"ugel 69, 53121 Bonn, Germany\\
              $^{2}$NRAO, PO Box O, Socorro, NM 87801, USA\\
              $^{3}$Raman Research Institute, CV Raman Ave,
              Sadashivanagar, Bangalore 560080, India\\
              $^{4}$Australia Telescope National Facility, PO Box 76,
              Epping 1710, NSW, Australia\\
              $^{5}$ASTRON, PO Box 2, 7990 AA, Dwingeloo, The Netherlands}
   \date{Received ; accepted}

   \abstract{ 
     {\it Context.}  Hydrogen ionized by young, high-mass stars in starburst
     galaxies radiates radio recombination lines (RRLs), whose strength can be
     used as a diagnostic of the ionization rate, conditions and gas dynamics
     in the starburst region, without problems of dust obscuration.  However,
     the lines are weak and only few extragalactic starburst systems have been
     detected.
     
     {\it Aims.} We aimed to increase the number of known starburst systems
     with detectable RRLs for detailed studies, and we used the line properties to
     study the gas properties and dynamics.
     
     {\it Methods.} We searched for the RRLs H91$\alpha$ and H92$\alpha$ with
     rest frequencies of 8.6 GHz and 8.3 GHz in the nearby southern Seyfert
     galaxy \object{NGC\,4945} using the Australia Telescope Compact Array with
     resolution of $3''$.  This yielded a detection from which we derived 
     conditions in the starburst regions.

     {\it Results.} We detected RRLs from the nucleus of NGC\,4945 with a peak
     line strength integrated over the source of $17.8$\,mJy, making it the
     strongest extragalactic RRL emitter known at this frequency.  The line
     and continuum emission from NGC\,4945 can be matched by a model
     consisting of a collection of 10 to 300 H\,II regions with temperatures of
     5\,000\,K, densities of $10^3$\,cm$^{-3}$ to $10^4$\,cm$^{-3}$ and a
     total effective diameter of 2\,pc to 100\,pc.  The Lyman continuum
     production rate required to maintain the ionization is $6 \times
     10^{52}~\mathrm{s}^{-1}$ to $3 \times 10^{53}~\mathrm{s}^{-1}$, which
     requires 2\,000 to 10\,000 O5 stars to be produced in the starburst,
     inferring a star formation rate of 2\,$M_\odot$\,yr$^{-1}$ to 
     8\,$M_\odot$\,yr$^{-1}$.  We
     resolved the rotation curve within the central 70\,pc region and this is
     well described by a set of rotating rings that were coplanar and edge on.
     We found no reason to depart from a simple flat rotation curve.  The
     rotation speed of 120\,km\,s$^{-1}$ within the central $1''$ (19 pc)
     radius infers an enclosed mass of $3 \times 10^7$\,$M_{\odot}$, and an
     average surface density with the central 19 pc of $25\,000$\,pc$^{-2}$,
     which exceeds the threshold gas surface density for star formation.
     
     {\it Conclusion.} We discovered RRLs from NGC\,4945.  It is the strongest
     known extragalactic RRL emitter and is suited to high-quality
     spectroscopic study.  We resolved the dynamics of the ionized gas in the
     central 70 pc and derived conditions and star formation rates in the
     ionized gas.

   }

   \maketitle
%

\section{Introduction}

NGC\,4945 is a nearby almost edge-on spiral galaxy at a distance of
(3.8\,$\pm$\,0.3)\,Mpc (\cite{karachentsev07}; implying a scale of 19\,pc per
arcsec).  It has prominent dust lanes obscuring the nucleus, and is one of the
brightest extragalactic sources seen by IRAS.  The nuclear optical spectrum
shows no sign of the Seyfert nucleus, and shows a purely starburst emission
(\cite{krabbe01} and references therein).  The presence of an active galactic
nucleus (AGN) was finally settled when X-ray emission from a variable compact
nuclear source was detected in the range 2\,keV to 10\,keV (\cite{iwasawa93})
with heavy absorption ($N_{\rm{H}} \sim 2\times10^{24}$\,cm$^{-2}$) typical of
Seyfert 2 galaxies.  The absorption was predicted to be much less at higher
X-ray energies, and indeed at 100\,keV \cite{done96} found it to be the
second-brightest known Seyfert galaxy in the sky.  X-ray imaging by
\cite{schurch02} with X-ray Multi-Mirror-Newton and Chandra revealed a conical
plume extending 500\,pc NW which they interpret as a mass-loaded superwind
driven by the starburst, and they found a 6.4\,keV iron K$\alpha$ line and a
Compton reflection component, which are characteristic of AGNs.

A compact ($< 30$\,mas) radio core with high brightness temperature
(10$^7$\,K) detected by \cite{sadler95} at 2.3\,GHz and 8.4\,GHz indicates the
presence of bright synchrotron emission, additional evidence for an AGN,
though could possibly be radio supernovae.  Mid-infrared spectroscopy was
carried out with the Infrared Space Observatory by \cite{spoon00} to penetrate
the dust obscuration, and they found no high excitation lines like [Ne\,V]
that are common in AGNs, implying huge obscuration of $A_{\rm{V}} > 160$
assuming that a narrow-line region were present like that seen in, for
example, the Circinus galaxy (which is similar in many ways to NGC\,4945).

The nucleus is the source of abundant infrared and molecular emission, and
optical emission line images show gas outflowing from the nucleus
perpendicular to the galaxy plane.  This activity originates from a composite
starburst and AGN.  An obscured nuclear starburst
ring with diameter 50 pc is seen in Pa$\alpha$ (\cite{marconi00}).

The molecular content of the NGC\,4945 nucleus is rich and varied, producing
the strongest known extragalactic molecular lines of many species.  Over 80
transitions from 19 molecules have been found with single dishes, principally
OH, H$_2$O, CO, HCN, HCO$^+$, CH$_3$OH, but also many others with the
Swedish-ESO-Submillimetre Telescope (SEST) and
Parkes from 1.6\,GHz to 354\,GHz (\cite{gardner74}, \cite{whiteoak74},
\cite{whiteoak80}, \cite{whiteoak90}, \cite{whiteoak86}, \cite{dosSantos79},
\cite{batchelor82}, \cite{henkel90}, \cite{henkel94}, \cite{curran01},
\cite{wang04}).  Many require densities of $10^5$\,cm$^{-3}$ for excitation
and yield cloud temperatures of $T_{\rm{kin}} \sim 100$\,K.  Wang et al. (2004) 
could determine isotope
ratios of C, N, O, and S, and see isotope enrichment by
ejecta from massive stars relative to that of fresh gas inflowing through a
bar showing that the starburst is old enough to have affected the isotopic
composition of the surrounding interstellar medium.  Interferometric imaging of 
the HNC and HCO$^+$ molecular distributions
by \cite{cunningham05} with the Australia Telescope Compact Array (ATCA) showed rotation of the molecular disk in
the central $6''$ with rotation speed of 135\,km\,s$^{-1}$.

Imaging with very long baseline interferometry of the H$_2$O masers by
\cite{greenhill97} resolved the rotation curve over the central 40 milliarcsec
(0.7\,pc diameter) with a velocity range of $\pm $150\,km\,s$^{-1}$ inferring
a central mass of $1.4 \times 10^6 M_\odot$.  The rotation is in the same
sense and in the same plane as the galaxy disk.

CO has been imaged at resolution from $43''$ to $15''$ with SEST by \cite{dahlem93}, 
\cite{ott01}, and \cite{mauersberger96}, and
interferometrically at $4''$ with the Submillimeter Array (SMA) by \cite{chou07}.  All groups
found the CO to be strongly concentrated towards the centre in a disk of
molecular material $16'' \times 11''$ (310\,pc $\times$ 210\,pc).  \cite{chou07} show
that the disk rotates, within the central $5''$ (95\,pc) radius in the plane of
the galaxy disk with simple rigid-body circular rotation. At larger radius the
isovelocity contours show an `S' shaped asymmetry due to a bar potential, and
at the centre there is an unresolved kinematically decoupled component with a
broad (340\,km\,s$^{-1}$) velocity range.  They find that this last component
is a good candidate for CO emission from the obscuring AGN torus, and too
dense to be part of the starburst-driven molecular outflow.

The ionized gas phase has been imaged in H$\alpha$ + N\,II, [O\,III], and
Br$\gamma$ by \cite{moorwood94}, \cite{moorwood96} and \cite{lipari97} and in
Pa$\alpha$ by \cite{marconi00}.  The optical lines show a conical wind-blown
cavity, and the infrared lines (Br$\gamma$ and Pa$\alpha$) show a starburst
disk at the base of the cavity extending over some $8''$ and so are embedded
within the molecular disk which extends to $16''$.  Despite the cone
resembling Seyfert ionization cones, it lacks [O\,III] emission and so that
origin has been excluded.  Spectra in H$\alpha$ + N\,II show motions of $\pm
500$ km s$^{-1}$ in the cone and optical line ratios typical of low-ionization
nuclear emission-line region galaxies.  Infrared spectra covering Br$\gamma$,
Pf$\beta$ and shocked molecular hydrogen by \cite{koornneef93},
\cite{moorwood94}, and \cite{spoon03} resolve the nuclear rotating disk over
the central $\pm 250$\,pc, showing a velocity range of 500\,km\,s$^{-1}$.  The
neutral hydrogen optical emission region is within the molecular hydrogen
region, with shock excitation from a nuclear wind from the central cluster.
The starburst seen in Br$\gamma$ is powerful enough to supply the $2 \times
10^{10} L_{\odot}$ radiated in the infrared.

The ionized gas component can also be imaged using radio recombination lines
(RRLs), which are not affected by extinction unlike at optical and
near-infrared wavelengths.  In this paper we describe RRL imaging of the
nucleus of NGC\,4945 in which we detect H92$\alpha$ emission and resolve the
rotation curve.  RRLs have been used to derive mass, density and ionizing
photon fluxes for the ionized gas in other galaxies, from which star formation
rates have been derived.  The diagnostic methods are described by
\cite{anantha00} and references therein, and in detail by Mohan
(2002).  The mere detection of RRL emission requires the presence of thousands
of H\,II regions or of stimulated emission since a single H\,II region like
the Orion nebula placed at that distance would be undetectably weak.

There are now 15 known extragalactic RRL detections (including NGC\,4945), all
in bright starburst galaxies, which are listed by \cite{roy08}.  Most of those
detections resulted from improving the search sensitivity by a factor of ten
during the 1990s and by making surveys of promising bright candidates using
the Very Large Array (VLA) and ATCA.  Our observation of NGC\,4945 reported here was the third
detection made in our ATCA survey for H92$\alpha$ emission.  The other two
detections from this survey (NGC\,3256 and the Circinus galaxy) were reported
by \cite{roy05} and \cite{roy08}).  NGC\,4945 has proven to be the strongest
known extragalactic RRL emitter on the sky.

We give velocities in the heliocentric frame using the optical velocity
definition throughout this paper.

\section{Observations} 

We observed NGC\,4945 with the ATCA simultaneously in the lines H91$\alpha$
and H92$\alpha$ with two orthogonal linear polarizations. The observing
parameters and results are summarized in Table 1.

Calibration and imaging were done using the AIPS software using standard
methods.  The flux-density scale assumed that PKS B1934-638 had a flux density
of 2.99\,~Jy at 8295\,MHz and 2.87\,Jy at 8570\,MHz, based on the \cite{baars77}
flux-density scale.  A phase calibrator was observed every half hour to
correct the instrumental phase response.  A bandpass calibrator was observed
every few hours for correcting the instrumental frequency response (bandpass).
Phase corrections obtained from self calibration of the continuum source were
applied to the spectral line data.  Continuum emission was subtracted from the
line data using the method UVLSF (\cite{cornwell92}) in which the
continuum is determined for each baseline by a linear fit to the spectrum.
The final continuum and line images were made using natural weighting of the
$(u, v)$ data to achieve maximum possible signal-to-noise ratio.

Uncertainties on the flux densities have an 11\,\% rms random multiplicative
component due to flux-density bootstrapping and atmospheric opacity, a
0.21\,mJy rms random additive component due to thermal noise in a 1\,MHz
channel and a systematic multiplicative component of 11\,\% rms due mainly to
the uncertainty in the Baars et al.  flux-density scale.


\begin{table*}
\caption[]{ATCA Observational Parameters and Results for NGC 4945.}
\label{ObsLine}
\scriptsize
\begin{center}
\begin{tabular}{lll}
\hline
\hline
\noalign{\smallskip}        & \multicolumn{1}{c}{NGC\,4945} \\
\noalign{\smallskip}
\hline
\noalign{\smallskip}

Array configuration         & 750D (1993jul25), 1.5D (1994sep21), 6C (94oct03) \\
                            
Date of observation         & 1993jul25, 1994sep21, 94oct03 \\

Integration time            & 22.7\,h \\

Transitions                 & H92$\alpha$ \& H91$\alpha$ \\

$\nu_{\rm rest}$ of RRL      & 8309.38\,MHz \& 8584.82\,MHz \\

$\nu_{\rm{band-centre}}$      & 8295.00\,MHz \& 8570.00\,MHz \\

$V_{\mathrm{systemic,helio,optical}}$
                             & 560\,km\,s$^{-1}$ (\cite{paglione97}) \\
                             
Distance                    & 3.8\,Mpc \\

Bandwidth, channels, IFs   &  64\,MHz, 64, 4 \\

Spectral resolution        &  36\,km\,s$^{-1}$ \\

Minimum baseline           & 0.86\,k$\lambda$ \\

Beam size (natural weight) &  $3.4'' \times 2.7'' $ at -55$^{\circ}$ \\

Beam size (uniform weight) &  $1.35'' \times 1.18'' $ at -4$^{\circ}$ \\

Phase calibrator (B1950)    & 1320-446 \\

Bandpass calibrator (B1950) & 1226+023, 1320-446, 1934-638, 2251+158 \\

\bf{Line and Continuum Properties} & \\

Peak continuum flux density  & $(538 \pm 80)$\,mJy\,beam$^{-1}$ (natural weighting)\\

Total continuum flux density & $(1424 \pm 220)$\,mJy  \\

Noise per image channel      & 0.21\,mJy\,beam$^{-1}$ \\

Peak line flux density       & $(17.8 \pm 2.8)$\,mJy \\
      
Integrated line flux         & $(11.3 \pm 1.8) \times 10^{-22}$\,W\,m$^{-2}$ \\

Measured line width (FWHM)   & $(225 \pm 15)$\,km\,s$^{-1}$   \\

Deconvolved line width (FWHM)  & $(221 \pm 15)$\,km\,s$^{-1}$  \\

Centroid helio. optical vel.   & $(581 \pm 15)$\,km\,s$^{-1}$   \\


No. of beam areas where line is observed     & 1.5 (naturally weighted)\\

\noalign{\smallskip}
\hline
\end{tabular}
\end{center}

\end{table*}



\section{Results}

The ATCA continuum and line images, and integrated spectrum are shown in Fig.
1, and the velocity field and position-velocity diagram are shown in Fig. 2.

\begin{figure}
\includegraphics[width=7cm]{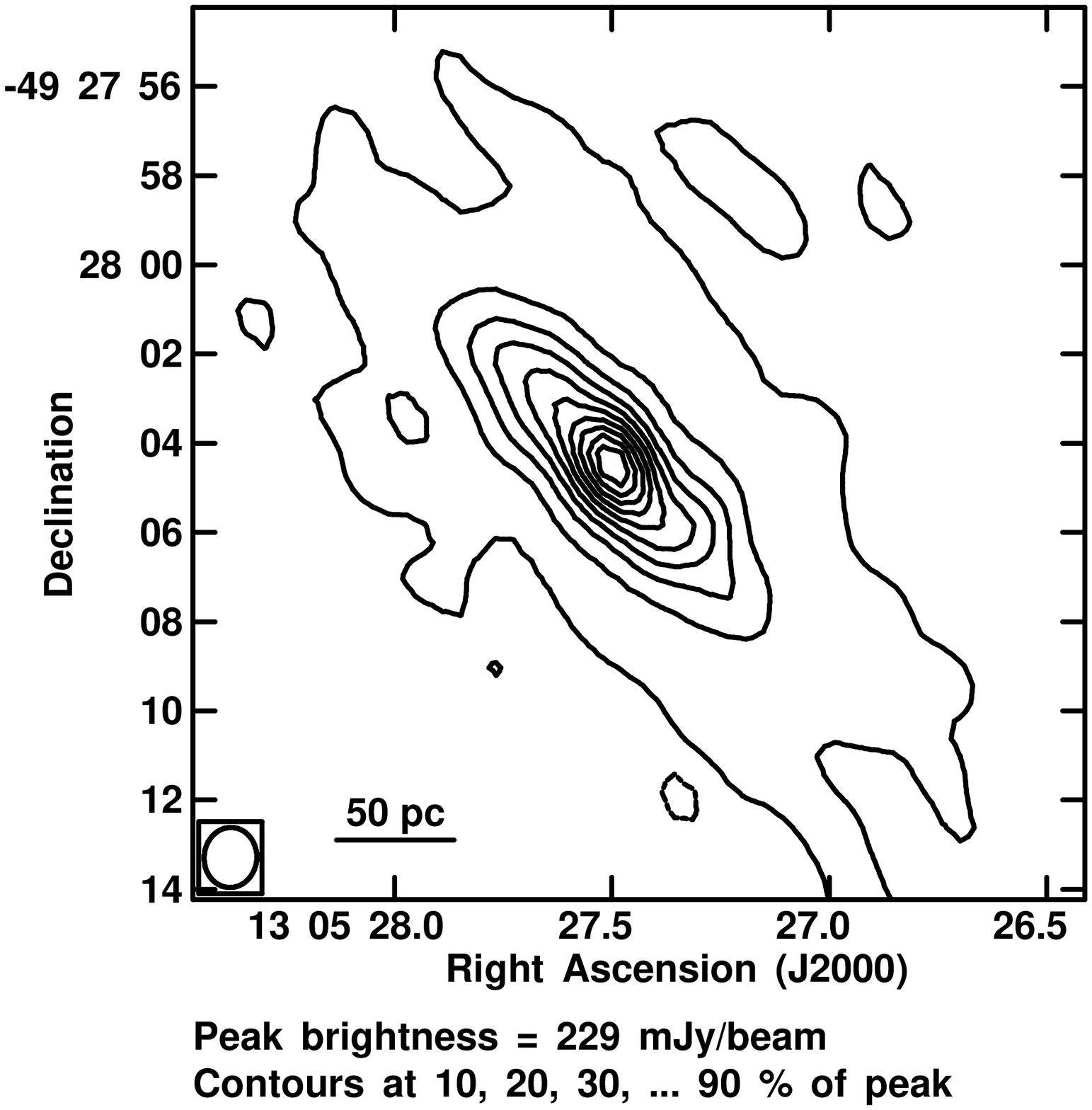}
\includegraphics[width=7cm]{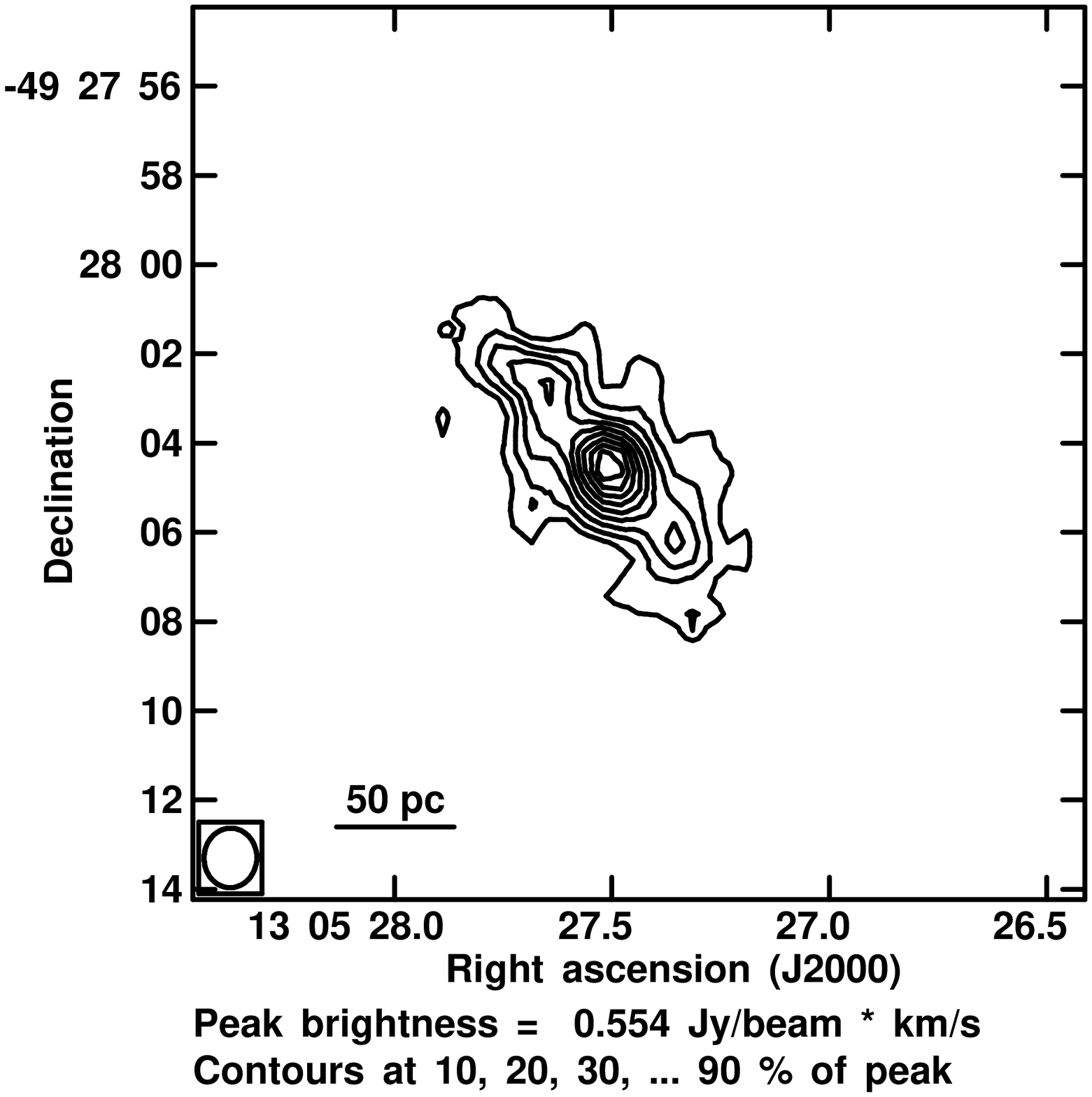}
\includegraphics[width=7cm]{13926fig1c.eps}
\caption{
Top: ATCA 8.3 GHz + 8.6 GHz continuum image with uniform weight
of NGC 4945.  
Middle: ATCA uniformly weighted zeroth-moment image of 
H91$\alpha$ + H92$\alpha$ emission after continuum subtraction. 
Beamsize is $1.4'' \times 1.2''$.
Bottom: ATCA H91$\alpha$ + H92$\alpha$ line profile integrated over the
line-emitting region in the zeroth-moment image. RMS noise is
0.16\,mJy\,beam$^{-1}$\,channel$^{-1}$ and the channel width is 
36\,km\,s$^{-1}$. The region of integration is 
a box of size $10.0'' \times 10.4''$ centred on RA 13 05 27.48 
dec -49 28 04.0 }
\end{figure}

The continuum image shows a well resolved structure extended along the plane
of the galaxy in position angle $47^{\circ}$.  Line emission was detected from
the nuclear region with a well resolved structure that is extended along the
plane of the galaxy like the continuum, though more compact and clumpy than
the continuum emission.  The line emitting region has a total deconvolved
extent of $8.1'' \times 1.7''$ (150\,pc $\times$ 30\,pc). The peak of line
emission is coincident with the peak of continuum emission within $1''$.  The
total area of line emission is 1.5 times the naturally-weighted beam area (8.6
times the uniformly-weighted beam area), or 4500\,pc$^{2}$.  The H91$\alpha$ +
H92$\alpha$ spectrum integrated over the line emitting region shows a strong
line detection with significance of 52\,$\sigma$, with centroid at
581\,km\,s$^{-1}$ in good agreement with the CO systemic velocity of
585\,km\,s$^{-1}$ (\cite{chou07}) and with the velocity of the HI
absorption towards the nucleus of 585\,km\,s$^{-1}$, though significantly
higher than the HI systemic velocity inferred from the whole galaxy of
557\,km\,s$^{-1}$ (\cite{ott01}).  The line FWHM is 220\,km\,s$^{-1}$ after
deconvolving the instrumental velocity resolution of 42\,km\,s$^{-1}$.

\begin{figure}
\includegraphics[width=7cm]{13926fig2a.eps}
\includegraphics[width=7cm]{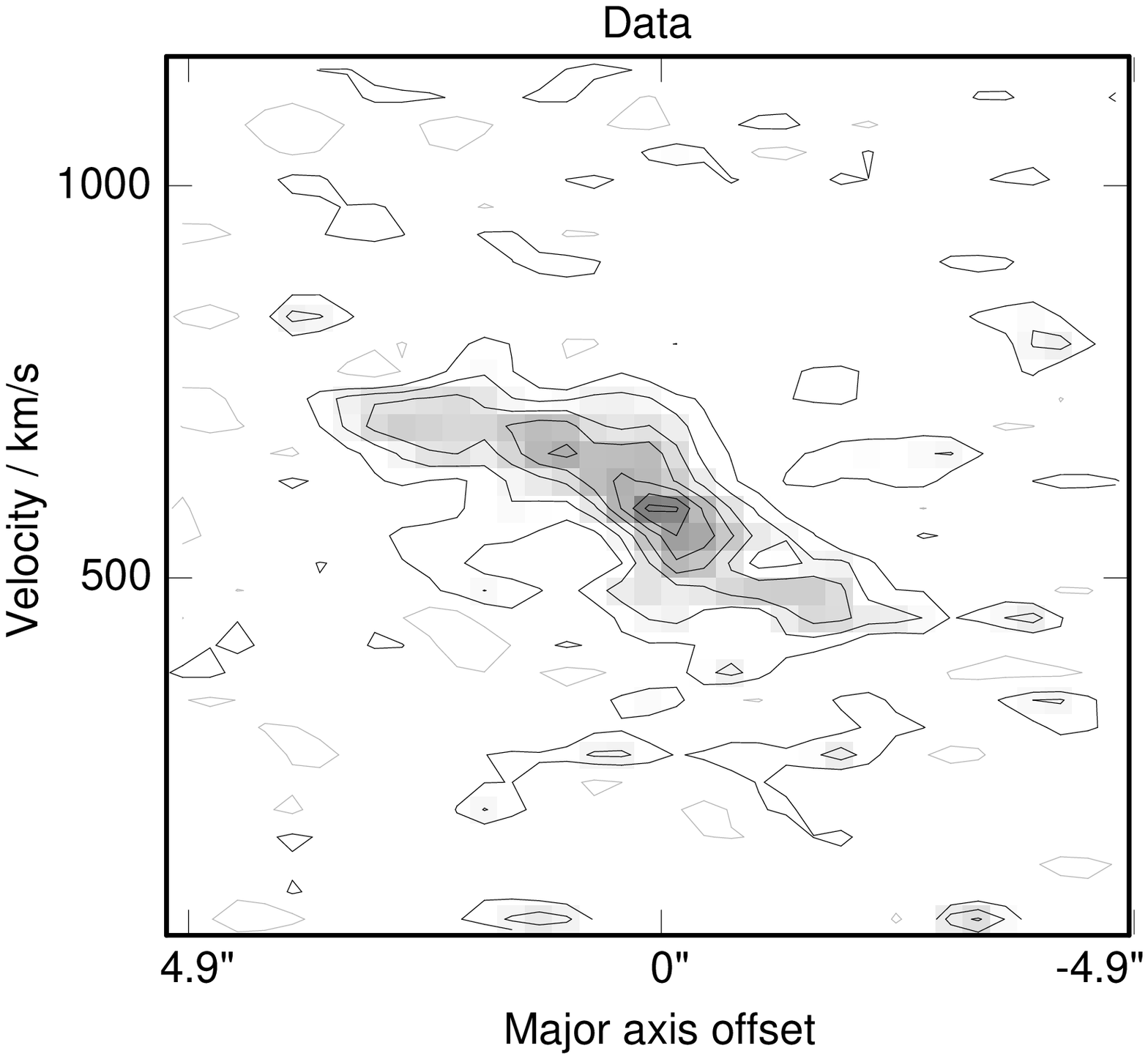}
\includegraphics[width=7cm]{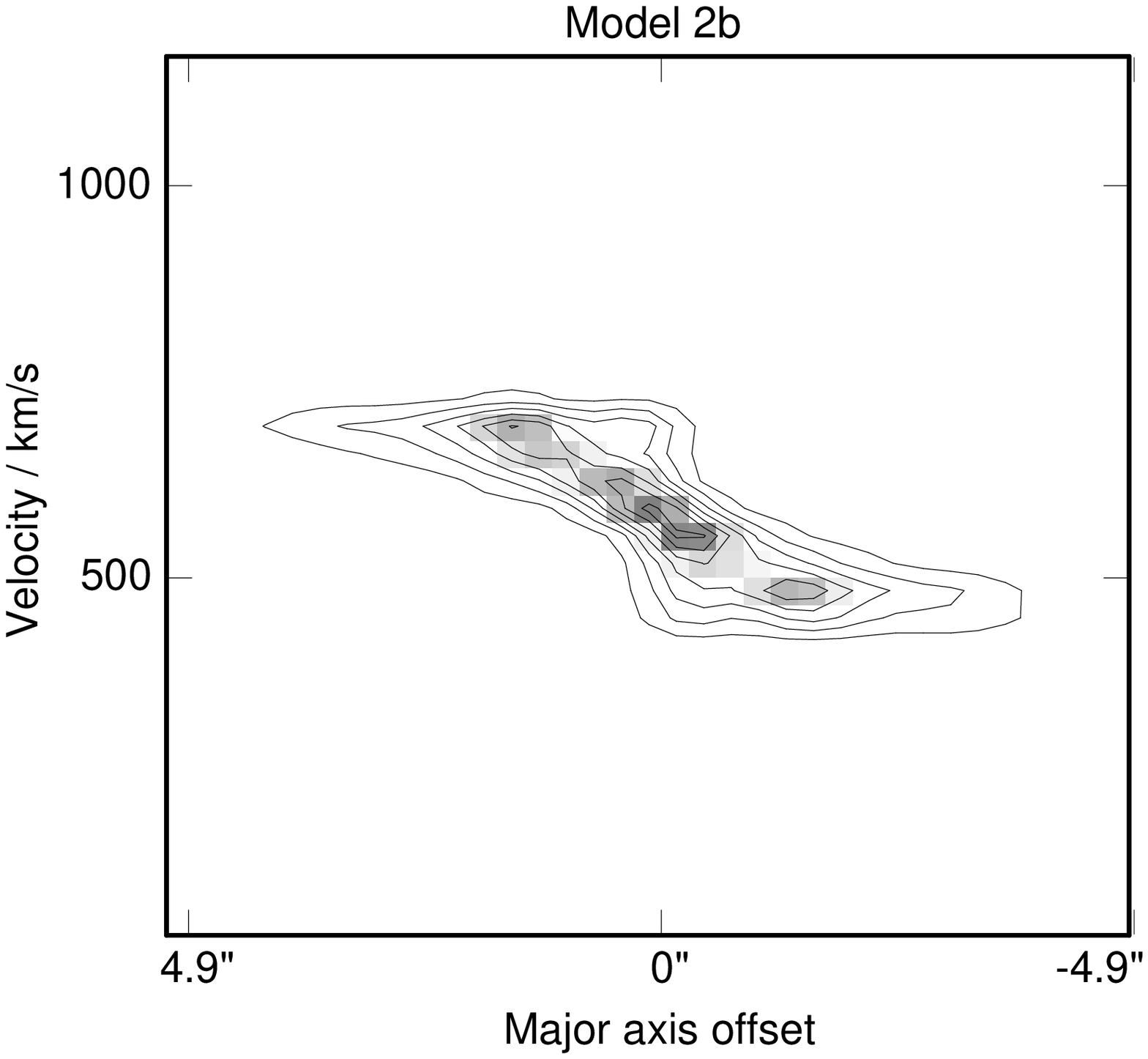}
\caption{
Top: ATCA first-moment image of H91$\alpha$ + H92$\alpha$ emission
showing rotation.  Beamsize is $1.4'' \times 1.2''$.
Middle: H91$\alpha$ + H92$\alpha$ rotation curve along the major axis
of the line-emitting region (PA = $47^{\circ}$).  Resolution is
  $1.3'' \times 36$\,km\,s$^{-1} $.
Bottom: Rotation curve from the best-fit dynamical model described in Sec. 5.
}
\end{figure}

The position-velocity diagram (Fig. 2) shows a well resolved symmetric structure
indicating undisturbed rotation with a flat rotation curve extending in 
close to the nucleus.  The rotation curve is modelled in Sec 5.

\section{Modelling Line Formation in the Ionized Gas}

Two types of models have been considered for the RRL emission from
the nuclei of external galaxies: one based on a uniform slab of ionized gas 
and the other based on a collection of compact H~II regions.  Such models
have been discussed by \cite{anantha00} and references therein,
and are documented in detail by \cite{mohan02}.
These models take as constraints the integrated RRL strength at one or more
frequencies, the radio continuum spectrum, and the geometry of the line
emitting region.

For modelling the RRL emission from NGC 4945 we used the integrated line
strength (17.8~mJy), line width (280~km~s$^{-1}$), size of the line-emitting
region (equivalent to a 70\,pc diameter sphere), continuum emission (1424 mJy), 
and spectral index (-0.75) to
constrain conditions in the ionized gas.  Using the collection of H\,II regions
model, models with 10 to 300 H~II regions, all with $T_{\mathrm{e}} \sim
5000$~K, $n_{\mathrm{e}} \sim 10^{3}$~cm$^{-3}$ to $10^{4}$~cm$^{-3}$ and a
total effective diameter of the line-emitting gas of 2\,pc to 100\,pc produced
good matches to the line and continuum emission.
Parameters derived for typical allowed models are given in Table 2.

\begin{table*}
\caption[]{Derived properties for NGC 4945, using model results for $T_{\rm e}
 = 5\,000$\,K}
\label{ModelRes1}
\scriptsize
\begin{center}
\begin{tabular}{llllllllll}
\hline
\hline
\noalign{\smallskip}
                            & NGC 4945 \\
\noalign{\smallskip}
\hline
\noalign{\smallskip}
Electron temperature        & 5\,000 K \\

Electron density            & 1000\,cm$^{-3}$ to 10\,000\,cm$^{-3}$ \\

Effective size $^a$         & 2\,pc to 100\,pc \\

Total ionized gas mass      & (2 to 6) $\times 10^{5}~M_{\odot}$ \\

$N_{\mathrm{LyC}}$          & $6 \times 10^{52}$~s$^{-1}$ to
                              $3 \times 10^{53}$~s$^{-1}$ \\

No. O5 stars                & 2\,000 to 10\,000 \\

\noalign{\smallskip}
\hline
\end{tabular}
\end{center}
$^{a}$ The total size, ie (total volume of all H\,II regions)$^{1/3}$ \\
\end{table*}

The range of possible filling factors can be estimated by comparing the
total volume of the line-emitting gas (8\,pc$^3$ to $1 \times 10^6$\,pc$^3$
derived from the total effective diameter) to the volume of the
line-emitting region ($5.3 \times 10^5$\,pc$^3$, assuming cylindrical
geometry with diameter of 150\,pc and height of 30\,pc).  The range of
possible values is then $2 \times 10^{-5}$ to 1.

The inferred mass of ionized gas is $2\times10^{5}~M_{\odot}$ to
$6\times10^{5}~M_{\odot}$, depending on the model conditions, which requires a
Lyman continuum flux of $6\times10^{52}$~s$^{-1}$ to $3\times10^{53}$~s$^{-1}$
to maintain the ionization.  This flux is equivalent to the Lyman continuum
output of 2000 to 10000 stars of type O5, which infers a star-formation rate
of 2\,$M_\odot$\,yr$^{-1}$ to 8\,$M_\odot$\,yr$^{-1}$ when averaged over the lifetime of OB stars.

This can be compared to star formation rates derived from other indicators
following \cite{hopkins03}.  Taking the peak 1.4\,GHz flux density of
4.0\,Jy\,beam$^{-1}$ in the $30''\times 18''$ beam of the ATCA at the nucleus
of NGC\,4945 (\cite{elmouttie97}) yields a 1.4 GHz luminosity of
$7.0\times10^{21}$\,W\,m$^{-2}$ and a corresponding star formation rate of
3.9\,$M_\odot$\,yr$^{-1}$.  This can be attributed to the central $1.8''$ (30 pc)
diameter region following the argument of \cite{elmouttie97} based on
the high resolution 8.4\,GHz image.  The IRAS $60\,\mu$m and $100\,\mu$m flux
densities yield a far-infrared (FIR) star formation rate of
4.1\,$M_\odot$\,yr$^{-1}$.  These estimates agree well with the star formation
rate estimated from the RRL emission of 2\,$M_\odot$\,yr$^{-1}$ to 8\,$M_\odot$\,yr$^{-1}$.  The
H$\alpha$ and U-band-based star formation rates were not estimated due to the
extreme absorption in the optical band.
The supernova rate estimated by \cite{forbes98} from the 4.8\,GHz flux density
is 0.23\,yr$^{-1}$, and from the [Fe II] flux is much lower at
0.005\,yr$^{-1}$.  These can be converted into star-formation rates following
\cite{fukugita03}, yielding rates of 19\,$M_\odot$\,yr$^{-1}$ and
0.4\,$M_\odot$\,yr$^{-1}$.  These bracket the range we estimated
from the RRL emission.

Our RRL-based SFR estimate assumed no dust absorption between the OB stars and
the H\,II regions producing the RRL emission.  The effect of dust can be
estimated by comparing the bolometric luminosity ($L_{\mathrm{bol-RRL}}$) from
the stellar population needed to ionize the RRL-emitting gas to the observed
FIR output ($L_{\mathrm{FIR}}$) from the region.  If ionizing photons are
absorbed by dust before causing ionization, then $L_{\mathrm{bol-RRL}}$ will
be less than $L_{\mathrm{FIR}}$.  The ratio $L_{\mathrm{bol-RRL}}$ /
$L_{\mathrm{FIR}}$ thus gives the ratio of the number of photons absorbed by
the gas to the number of photons absorbed by the dust.  We derived
$L_{\mathrm{bol-RRL}}$ by taking our RRL-based $L_{\mathrm{LyC}}$ and dividing
by the $L_{\mathrm{LyC}}$ / $L_{\mathrm{bol}}$ ratio of 0.29 calculated for a
stellar population with Salpeter IMF, lower-mass cutoff of $1\,M_{\odot}$ and
upper-mass cutoff of $87\,M_{\odot}$, and using the mass-luminosity relation
for OB stars in Table 5 of \cite{vacca96}.  This resulted in
$L_{\mathrm{bol-RRL}} = 1.2 \times 10^9 L_{\odot}$ to $5.7 \times 10^9
L_{\odot}$, compared to $L_{\mathrm{FIR}}$ from region of $1.6 \times 10^{10}
L_{\odot}$ (Brock et al. 1988).  The ratio $L_{\mathrm{bol-RRL}}$ /
$L_{\mathrm{FIR}}$ shows that the gas volume in the nuclear region is illuminated
by ionizing radiation that represents only 7.5\,\% to 35\,\% of
the ionizing flux from the stars needed to power the FIR output; the
rest, we assume, is absorbed by dust within the H\,II regions.  Converting to
opacity using Fig 2 of \cite{petrosian72} yields dust opacities of 1.3
neper to 3.4 neper within the H\,II regions.

Given so much absorption, it is unexpected that the star-formation
rate inferred from the RRL emission, which assumed zero absorption,
should agree so well with that inferred from $L_{\mathrm{FIR}}$.  An important
assumption made while deriving the RRL-based star formation rate from
the present inferred number of OB stars was that the stars are being
formed at a steady rate that would maintain the present numbers.

\section{Dynamical Modelling of the Ionized Gas}

We used the rotation curve to constrain the gas kinematics by fitting
to the data a simple model consisting of a set of rings, coplanar, edge-on,
with an initially flat rotation curve.  The brightness of each ring was
determined by deprojecting the observed zeroth-moment image
to derive the radial distribution of the line intensity.
The radial distribution showed a central peak and a ring of
emission $2.5''$ (50 pc) from the nucleus.

We refined the model iteratively, varying the systemic velocity, rotation
velocity, and velocity dispersion until we achieved a reasonably close match
to the data, and found no reason to depart from a simple flat rotation curve.
The biggest residual between the model and data was at the nucleus and a
smaller one at the 50~pc ring.  To improve the model, we refined the radial
profile of line emission strength at the nucleus and the ring.  The data
required us to boost the central peak to 25 times the brightness of the more
extended emission and to place that gas at the systemic velocity.  This could
correspond to gas on circular orbits moving transverse to the line of sight in
front of the nucleus.  The data also required us to put half that strength in
the innermost ring (0.4" = 7.6 pc) with our best-fit rotation velocity.  For
the 50~pc ring, we tweaked the radius and brightness.  The result was a very
good match to the observed data, with residuals at 1.5 times the rms noise in
the position-velocity diagram.

The final model had a flat rotation
curve with $v_{\rm{systemic}} = 581$~km~s$^{-1}$, $v_{\rm{rotation}}$ =
120~km~s$^{-1}$, and $v_{\rm{dispersion}}$ = 15~km~s$^{-1}$.  We did not need
to invoke a bar or radial motion, though the data do not strongly exclude
either.

The rotation velocity of 120~km~s$^{-1}$ within the central $1''$ (19~pc) radius
infers an enclosed mass of $3 \times 10^7 M_\odot$.  The water masers
(\cite{greenhill97}) infer an enclosed mass of $1 \times 10^6 M_\odot$
within 0.3~pc radius, and so most of the $3 \times 10^7 M_\odot$ is extended
between 0.3\,pc and 19\,pc radius of the nucleus.  The average surface density
within the central 19\,pc is $25\,000~M_\odot~$pc$^{-2}$, which
exceeds the threshold gas surface density for star-formation of (3 to
10)~$M_\odot~$pc$^{-2}$ (\cite{kennicutt89}) by four orders of magnitude.

The 50~pc ring might be a Lindblad resonance, which would infer a bar and then
our assumption of pure rotation would become invalid.  The ring and nucleus
may be seen also in CO spectra (\cite{whiteoak90}, \cite{dahlem93}), and
in OH absorption (\cite{whiteoak90}).

The RRL emission contains a bright peak on the nucleus at the systemic
velocity.  The continuum emission is smooth and does not show a similar peak
and so we speculate that this might be stimulated emission along the line of
sight for which the gas velocity is transverse to the line of sight and so 
has a long maser gain path.

\section{Comparison with literature}

H\,I kinematics were studied by \cite{ott01} with the ATCA at $3.5''$
resolution.  They saw HI in emission over the disk, with an extent of $22'
\times 4'$ (43\,kpc x 7.8\,kpc) and in absorption towards the nuclear
continuum emission.  They see an asymmetry in the large-scale H\,I rotation
field, which they argue is due to the bar potential.  The H\,I absorption
towards the nucleus shows a uniform velocity gradient across the nucleus
indicating solid-body rotation peaking at $\pm 100$\,km\,s$^{-1}$ either side
of the systemic velocity out to $\pm 5''$ either side of the centre, with a
sign of flattening beyond $\pm 5''$ to $\pm 7''$.  For comparison, our
H91$\alpha$ + H92$\alpha$ rotation curve remained flat at $\pm
120$\,km\,s$^{-1}$ either side of systemic all the way into the central $1''$.  This
difference could be due to the two-times lower spatial resolution of the HI
observation compared to that of the H91$\alpha$ + H92$\alpha$ observation
($3.5''$ compared to $1.4''$).  Were we to smooth our position-velocity
diagram, or had the noise in our position-velocity diagram been a little
higher, the small wings that indicate a flat rotation curve would be less
visible, giving the impression of solid-body rotation.  Also, in edge-on
systems with a flat rotation curve, one finds the density distribution often
gives the appearance of solid-body rotation.

Rotation of the larger-scale disk ($> \pm 100''$) derived from CO(1-0)
observations with SEST by \cite{dahlem93} at $43''$ resolution shows evidence
for gas inflow that is consistent with the inflow model proposed by
\cite{ables87} from H\,I.  This was seen also in CO(2-1) by \cite{ott01} with
SEST at $23''$ resolution.  However, inflow was not seen in the H91$\alpha$ +
H92$\alpha$ dynamics within $4''$ of the nucleus.

Isovelocity contours of CO by \cite{chou07} show an S-shaped asymmetry
due to a bar potential at distances larger than $5''$ from the nucleus, and
show simple circular rotation within that radius.  The isovelocity contours 
of the H91$\alpha$ + H92$\alpha$ radio recombination line from the ionized 
gas component seen within $4''$ radius of the nucleus likewise shows
simple circular rotation with no asymmetry, consistent with the CO on that
scale.

The rotation of the nuclear disk derived from CO(3-2)
observations with SEST by \cite{mauersberger96} with $15''$ resolution and with
SMA by \cite{chou07} at $\sim 4''$ resolution show that within $\pm 5''$ of
the nucleus, the velocity changes from 390\,km\,s$^{-1}$ to 730\,km\,s$^{-1}$,
which is broader than the velocity range of the H91$\alpha$ + H92$\alpha$ emission
(460\,km\,s$^{-1}$ to 700\,km\,s$^{-1}$) over the same region.

The CO emission seen by both \cite{mauersberger96} and \cite{chou07} shows two
local intensity peaks, with a radius of $6''$ to $8''$ from the nucleus.  The
H91$\alpha$ + H92$\alpha$ line emission radial distribution showed a central
peak and a ring of emission with radius $2.5''$ (50 pc) from the nucleus.
Thus, the ring in ionized gas lies within the ring in the molecular gas,
consistent with the illustration by \cite{spoon03} Fig. 8.

The dynamical mass within the central 100\,pc radius derived from the CO by
\cite{mauersberger96} is $8 \times 10^8 M_{\odot}$ and the dynamical mass
within 0.3\,pc radius derived from the water masers by \cite{greenhill97} is
$1.0 \times 10^6 M_\odot$.  The dynamical mass enclosed within the central
$1''$ (19~pc) radius derived from the H91$\alpha$ + H92$\alpha$ line is $3
\times 10^7 M_\odot$, which lies between the masses within the larger and
smaller radii, as expected.

Rotation of the nuclear ionized gas is traced also by Pf$\beta$ by
\cite{spoon03}.  They find ionized gas within the central $\pm 5''$ around the
nucleus from the same region in which we find H91$\alpha$ + H92$\alpha$
emission.  Comparing their position-velocity diagram of Pf$\beta$ (their
Fig. 7) to our position-velocity diagram of H91$\alpha$ + H92$\alpha$ (our
Fig. 2), the diagrams look essentially the same, having the same central
velocity gradient, the same rotation rate, the same systemic velocity, and the
same overall extent.  Thus, both emission lines are tracing the same gas, as
expected.  \cite{spoon03} propose a geometry in which the HI Pf$\beta$
originates from the central ionized and rotating disk (their Fig. 8).  The
H91$\alpha$ + H92$\alpha$ kinematics supports this picture.  Similar rotation
curves are seen in HCN, HNC, and HCO$^+$ by \cite{cunningham05}, though the
molecular material is seen more strongly concentrated into an edge-on ring
with weak emission or absorption towards the nucleus, unlike the ionized gas
which is always in emission.

\cite{chou07} identify a kinematically decoupled component in their CO
position-velocity diagrams along the minor axis, showing broad velocity width
at or near the disk centre, as a strip in velocity at the origin of the
position-velocity diagram spanning 400\,km\,s$^{-1}$ to 750\,km\,s$^{-1}$ in
their Fig. 5.  They identify this gas as a promising candidate for the
circumnuclear molecular torus invoked by AGN unification models.  For
comparison, we show the position-velocity diagram in H91$\alpha$ + H92$\alpha$
along the minor axis in Fig. 3.  The H91$\alpha$ + H92$\alpha$ shows a broad
velocity range at the position of the nucleus (panel 252), spanning
430\,km\,s$^{-1}$ to 750\,km\,s$^{-1}$, similar to that seen in CO by
\cite{chou07}.  This broad velocity width in the centre of our
position-velocity diagram along the minor axis underlines that the rotation
curve is flat; at this position close to the centre, one sees the full
receding and the full approaching velocities, which can be the case only if
the rotation curve rises very quickly in the central beam and stays flat
further out, as seen the other panels of Fig 3.  The gas with broad velocity
width at the nucleus was included when we fit the kinematics and we obtained a
good fit with a simple flat rotation curve extending into the central ring; we
did not need to introduce any departure from the simple flat-rotation-curve
model.

\onecolumn
\begin{figure}
\includegraphics[width=16cm]{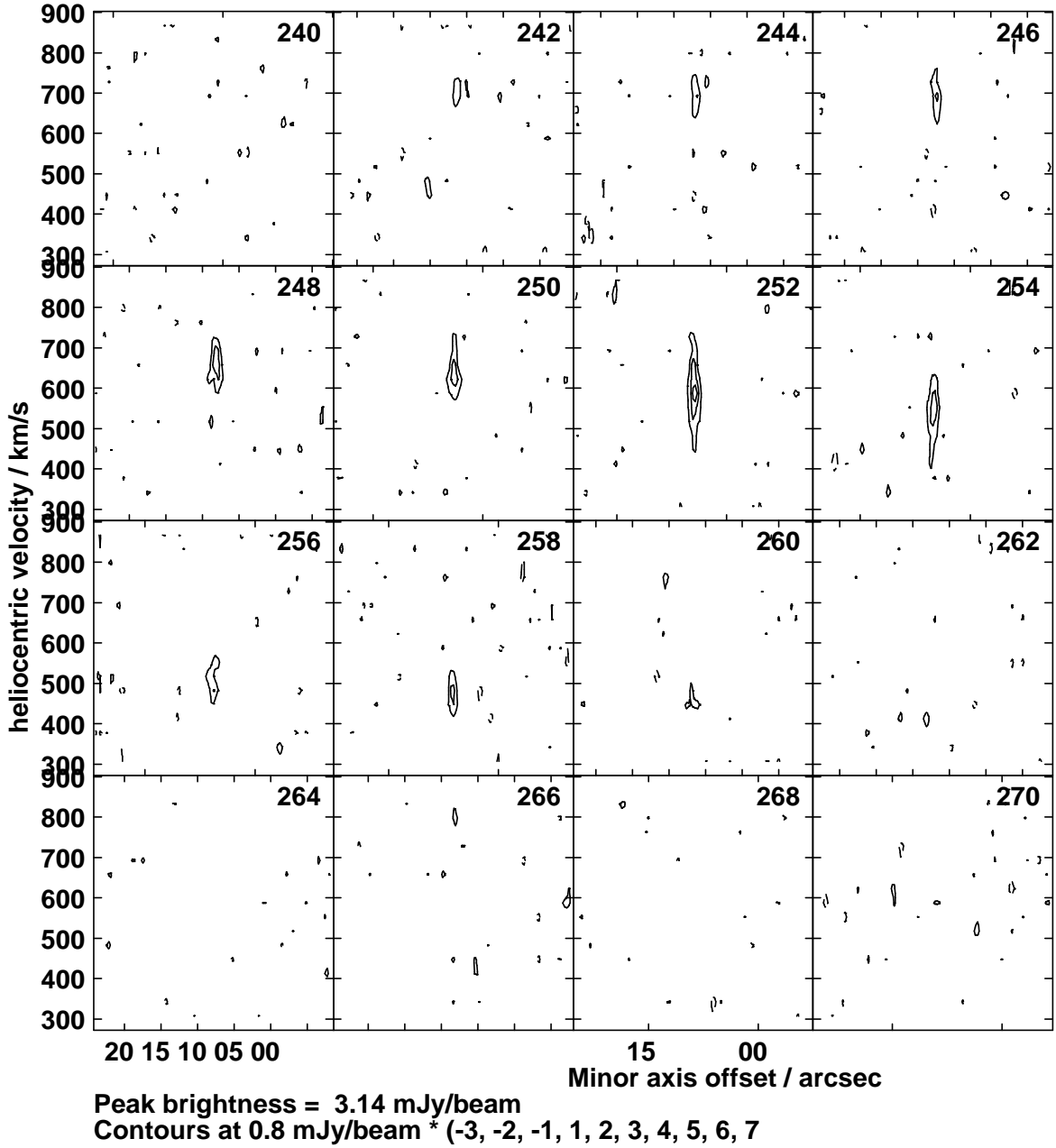}
\caption{
  H91$\alpha$ + H92$\alpha$ rotation curves along the minor axis direction of
  the line-emitting region (PA = $-43^{\circ}$).  Each panel is separated by
  $0.8''$ along the major axis, with the slice shown in panel 252 passing 
  through the peak of emission at the nucleus.  The minor axis offset 
  values are relative to an arbitrary offset.  Resolution is
  $1.3'' \times 36$\,km\,s$^{-1} $.  
}
\end{figure}
\twocolumn

\section{Conclusions}

We have discovered H91$\alpha$ + H92$\alpha$ lines in emission in NGC\,4945
with flux density of 17.8\,mJy using the ATCA, making NGC\,4945 the brightest
known extragalactic RRL source.  The detected line strength infers an ionized
gas mass of (2 to 6)\,$\times\,10^{5}\,M_{\odot}$ and a corresponding star
formation rate of 2\,$M_\odot$\,yr$^{-1}$ to 8\,$M_\odot$\,yr$^{-1}$ depending 
on the model
conditions.  The star formation rate estimated from the RRL detection in
NGC\,4945 agrees well with rates estimated from radio and FIR
luminosities using previously-calibrated relations.

The rotation curve was found to be flat into the central 1'' with 
$v_{\rm{systemic}} = 581$~km~s$^{-1}$, $v_{\rm{rotation}}$ =
120~km~s$^{-1}$.  We found no need to invoke a bar or radial motion and no
indication of a kinematically decoupled component.

Future observations at high frequencies where RRLs are stronger and
resolution is higher will provide measurements of multiple
transitions to provide better constraints on the gas conditions.

Since RRLs occur over a wide range of wavelengths, a given array can provide
higher resolution for studying the dynamics of ionized gas than can be
achieved for studies of neutral hydrogen.

Followup with the Atacama Large Millimeter/submillimeter Array of RRLs in
NGC~4945 and others should be rewarding, giving good SNR and high spatial
resolution to see finer details in the kinematics and of the possible
circumnuclear torus.

\begin{acknowledgements}

The Australia Telescope Compact Array is part of the
Australia Telescope, which is funded by the Commonwealth of Australia for
operation as a National Facility managed by CSIRO.

The National Radio Astronomy Observatory is a facility of the National Science
Foundation operated under cooperative agreement by Associated Universities,
Inc.

\end{acknowledgements}


\begin{thebibliography}{}

\bibitem[Ables et al. (1987)]{ables87} Ables, J. G., Forster, J. R.,
  Manchester, R. N., Rayner, P. T., Whiteoak, J. B., Mathewson, D. S.,
  Kalnajs, A. J., Peters, W. L., \& Wehner, H. 1987, MNRAS, 226, 157

\bibitem[Anantharamaiah et al. (2000)]{anantha00} Anantharamaiah, K. R.,
  Viallefond, F., Mohan, N. R., Goss, W. M., \& Zhao, J. H. 2000, ApJ, 537, 613

\bibitem[Baars et al. (1977)]{baars77} Baars J. W. M., Genzel, R.,
  Pauliny-Toth, I. I. K., Witzel, A.
  1977, A\&A, 61, 99
  
\bibitem[Batchelor et al. (1982)]{batchelor82} Batchelor, R. A., Jauncey,
  D. L., \& Whiteoak, J. B. 1982, MNRAS, 200, 19P

\bibitem[Chou et al. (2007)]{chou07} Chou, R. C. Y., Peck, A. B., Lim, J.,
  Matsushita, S., Muller, S., Sawada-Satoh, S., Dinh-V-Trung, Boone, F.,
  \& Henkel, C. 2007, ApJ, 670, 116

\bibitem[Cornwell, Uson \& Haddad (1992)]{cornwell92} Cornwell, T. J., Uson,
  J. M. \& Haddad, N. 1992, A\&A, 258, 583

\bibitem[Cunningham \& Whiteoak (2005)]{cunningham05} Cunningham, M. R., \&
  Whiteoak, J. B. 2005, MNRAS, 364, 37

\bibitem[Curran et al. (2001)]{curran01} Curran, S. J., Johansson, L. E. B.,
Bergman, P., Heikkil\"a, A., \& Aalto, S. 2001, A\&A, 367, 457

\bibitem[Dahlem et al. (1993)]{dahlem93} Dahlem, M., Golla, G., Whiteoak,
  J. B., Wielebinski, R., H\"uttemeister, S. \& Henkel, C. 1993, A\&A, 270, 29

\bibitem[Done et al. (1996)]{done96} Done, C., Madejski, G.~M. \& Smith, D.~A. 1996,
ApJ, 463, L63

\bibitem[Dos Santos \& L\'epine (1979)]{dosSantos79} Dos Santos, P. M., \&
  L\'epine, J. R. D. 1979, Nature, 278, 34

\bibitem[Elmouttie et al. (1997)]{elmouttie97} Elmouttie, M., Haynes, R. F.,
\& Jones, K. L., Ehle, M., Beck, R., Harnett, J. I., \& Wielebinski, R. 
1997, MNRAS, 284, 830

\bibitem[Forbes \& Norris (1998)]{forbes98} Forbes, D. A., \& Norris,
  R. P. 1998, MNRAS, 300, 757

\bibitem[Fukugita \& Kawasaki (2003)]{fukugita03} Fukugita, M., \& Kawasaki,
  M. 2003, MNRAS, 340, L7

\bibitem[Gardner \& Whiteoak (1974)]{gardner74} Gardner, F. F., \& Whiteoak,
  J. B. 1974, Nature, 247, 526

\bibitem[Greenhill et al. (1997)]{greenhill97} Greenhill, L.~J., Moran, J.~M. \& 
Herrnstein, J.~R. 1997, ApJ, 481, L23

\bibitem[Henkel et al. (1990)]{henkel90} Henkel, C., Whiteoak, J. B., Nyman,
L.-\AA., \& Harju, J. 1990, A\&A, 230, L5

\bibitem[Henkel et al. (1994)]{henkel94} Henkel, C., Whiteoak, J. B.,
\& Mauersberger, R. 1994, A\&A, 284, 17

\bibitem[Hopkins et al. (2003)]{hopkins03} Hopkins, A. M., Miller, C. J.,
Nichol, R. C., Connolly, A. J., Bernardi, M., G\'omez, P. L., Goto, T.,
Tremonti, C. A., Brinkmann, J., Ivezi\'c, \'Z., \& Lamb, D. Q. 2003, ApJ, 599, 971

\bibitem[Iwasawa et al. (1993)]{iwasawa93} Iwasawa, K., Koyama, K., Awaki, H.,
  Hunieda, H., Makishima, K., Tsuru, T., Ohashi, T., \& Nakai, N. 1993, ApJ,
  409, 155

\bibitem[Karachentsev et al. (2007)]{karachentsev07} Karachentsev, I. D.,
  Tully, R. B., Dolphin, A., Sharina, M., Makarova, L., Makarov, D., Sakai,
  S., Shaya, E. J., Kashibadze, O. G., Karachentseva, V., \& Rizzi, L. 2007,
  AJ, 133, 504

\bibitem[Kennicutt (1989)]{kennicutt89} Kennicutt, R.~C. 1989, ApJ, 344, 685
  
\bibitem[Koornneef (1993)]{koornneef93} Koornneef, J. 1993, ApJ, 403, 581

\bibitem[Krabbe et al. (2001)]{krabbe01} Krabbe, A., B\"oker, T., \& Maiolino,
  R. 2001, ApJ, 557, 626

\bibitem[L\'{i}pari et al. (1997)]{lipari97} L\'ipari, S., Tsvetanov, Z., \&
  Macchetto, F. 1997, ApJS, 111, 369

\bibitem[Marconi et al. (2000)]{marconi00} Marconi, A., Oliva, E., van der
  Werf, P. P., Maiolino, R., Schreier, E. J., Macchetto, F., \& Moorwood
  A. F. M. 2000, A\&A, 357, 24

\bibitem[Mauersberger et al. (1996)]{mauersberger96} Mauersberger, R., Henkel,
  C., Whiteoak, J. B., Chin, Y.-N., \& Tieftrunk, A. R. 1996, A\&A, 309, 705

\bibitem[Mohan (2002)]{mohan02} Mohan, Niruj R. 2002, PhD Thesis, IISc,
  Bangalore, India

\bibitem[Moorwood \& Oliva (1994)]{moorwood94} Moorwood, A. F. M., \& Oliva,
  E. 1994, ApJ, 429, 602

\bibitem[Moorwood et al. (1996)]{moorwood96} Moorwood, A. F. M., van der Werf,
  P. P., Kotilainen, J. K., Marconi, A., \& Oliva, E. 1996, A\&A, 308, L1

\bibitem[Ott et al. (2001)]{ott01} Ott, M., Whiteoak, J. B., Henkel, C., \&
  Wielebinski, R. 2001, A\&A, 372, 463

\bibitem[Paglione (1997)]{paglione97} Paglione, T. A. D., Jackson, J. M., 
Ishizuki, S. 1997, ApJ, 484, 656

\bibitem[Petrosian (1972)]{petrosian72} Petrosian, V. Silk, J. \& Field,
  G. B. 1972, ApJ, 177, L69

\bibitem[Roy et al. (2005)]{roy05} Roy, A. L., Goss, W. M., Mohan, Niruj R., 
Anantharamaiah, K. R. 2005, A\&A, 435, 831

\bibitem[Roy et al. (2008)]{roy08} Roy, A. L., Goss, W. M., 
Anantharamaiah, K. R. 2008, A\&A, 483, 79

\bibitem[Sadler et al. (1995)]{sadler95} Sadler, E. M., Slee, O. B., Reynolds,
  J. E., \& Roy, A. L. 1995, MNRAS, 276, 1373

\bibitem[Schurch et al. (2002)]{schurch02} Schurch, N. J., Roberts, T. P., \&
  Warwick, R. S. 2002, MNRAS, 335, 241

\bibitem[Spoon et al. (2000)]{spoon00} Spoon, H. W. W., Koornneef, J.,
  Moorwood, A. F. M., Lutz, D., \& Tielens, A. G. G. M. 2000, A\&A, 357, 898

\bibitem[Spoon et al. (2003)]{spoon03} Spoon, H. W. W., Moorwood, A. F. M.,
Pontoppidan, K. M., Cami, J., Kregel, M., Lutz, D., \& Tielens,
A. G. G. M. 2003, 
A\&A, 402, 499

\bibitem[Vacca et al. (1996)]{vacca96} Vacca, W. D., Garmany, C. D., \& Shull, 
J. M. 1996, ApJ, 460, 914

\bibitem[Wang et al. (2004)]{wang04} Wang, M., Henkel, C., Chin, Y. -N.,
  Whiteoak, J. B., Hunt Cunningham, M., Mauersberger, R., \& Muders, D. 2004,
  A\&A, 422, 883

\bibitem[Whiteoak \& Gardner (1974)]{whiteoak74} Whiteoak, J. B., \& Gardner,
  F. F. 1974, ApL, 15, 211

\bibitem[Whiteoak \& Gardner (1986)]{whiteoak86} Whiteoak, J.~B. \& Gardner, F.~F. 1986,
MNRAS, 222, 513

\bibitem[Whiteoak et al. (1980)]{whiteoak80} Whiteoak, J. B., Gardner, F. F.,
  \& Hoglund, B. 1980, MNRAS, 190, 17P

\bibitem[Whiteoak \& Wilson (1990)]{whiteoak90} Whiteoak, J. B., \& Wilson,
  W. E. 1990, MNRAS, 245, 665

\end{thebibliography}
\end{document}